\def\Wjjj{$W\,\!+\,3$}
\def\jets{\,{\rm jets}\,}

\documentclass{PoS}

\title{
\rule{0cm}{2.5cm}\vspace{-3.5cm}\\{ \it \normalsize IPPP-09-67 \hfill MIT-CTP-4074 \hfill Saclay--IPhT--T09/136  \\
SLAC--PUB--13794  \hfill UCLA/09/TEP/201 
 } \vspace{1cm}\\NLO QCD Predictions for $\boldmath{W+3}$ jets 
}

\ShortTitle{NLO QCD Predictions for $W+3$ jets}

\author{\speaker{Daniel Ma\^{\i}tre}
	\\
        Department of Physics, University of Durham,
          DH1 3LE, UK\\
        E-mail: \email{daniel.maitre@durham.ac.uk}}
\author{Carola F. Berger\\
        Center for Theoretical
Physics, Massachusetts Institute of Technology,
      Cambridge, MA 02139, USA\\
        E-mail: \email{cfberger@mit.edu}}
\author{Zvi Bern, Fernando Febres Cordero, Harald Ita\\
        Department of Physics and Astronomy, UCLA, Los Angeles, CA
90095-1547, USA\\
        E-mail: \email{bern@physics.ucla.edu}, \email{ffebres@physics.ucla.edu}, \email{ita@physics.ucla.edu}}
\author{Lance J. Dixon, Darren Forde, Tanju Gleisberg\\
        SLAC National Accelerator Laboratory, Stanford University,
             Stanford, CA 94309, USA\\
        E-mail: \email{lance@slac.stanford.edu}, \email{dforde@slac.stanford.edu}, \email{tanju@slac.stanford.edu}}
\author{David Kosower\\
        Institut de Physique Th\'eorique, CEA--Saclay,
          F--91191 Gif-sur-Yvette cedex, France\\
        E-mail: \email{david.kosower@cea.fr}}

\abstract{
In this contribution we present results from the NLO 
computation of the production of a $W$ boson in association 
with three jets in hadronic collisions.  The results are obtained by
combining two programs: {\tt BlackHat} for the virtual one-loop
matrix elements and {\tt Sherpa} for the real-emission contributions.
We present results for the Tevatron and the LHC, and address the issue
of the choice of a common factorization and renormalization scale
for this process.
}

\FullConference{European Physical Society Europhysics Conference on
High Energy Physics, EPS-HEP 2009,\\
July 16 - 22 2009\\ Krakow, Poland}

\begin{document}

\section{Introduction}

The production of a $W$ boson in association with jets forms a key 
set of processes at hadron colliders.  These processes are not just
important benchmarks in understanding the Standard Model at colliders;
they also constitute important backgrounds to top-quark production
and to new physics signals.  In addition, inclusive
$W$ production is a means for determining the (partonic) luminosity
at the LHC. An important aspect of such studies is to have good theoretical
control over Standard Model predictions.  Precise predictions are first
obtained at next to leading order (NLO) in the QCD coupling. In
contrast, leading-order (LO) computations usually suffer from large
renormalization and factorization scale uncertainties. More importantly,
shapes of distributions can be altered by higher-order corrections.

NLO predictions have been available for processes involving a $W$
boson and up to two jets \cite{MCFM}. The bottleneck for the
inclusion of a third jet had been the computation of the virtual
matrix elements.  Recently, significant progress has been made
on this problem~\cite{ICHEPBH,BergerW3}.  In
these proceedings, we summarize results from the first complete NLO
calculation of \Wjjj\jets~\cite{BergerW3}.

NLO results are obtained by combining the Born-level matrix elements
with two additional contributions, the virtual one-loop amplitude
interfered with the tree-level amplitude, and the squared real-emission
matrix elements. These two additional contributions are provided by
two different programs, {\tt BlackHat} for the one-loop matrix
elements and {\tt Sherpa} for the real-emission contribution.

{\tt Sherpa} \cite{SHERPA} is a Monte Carlo event generator framework
written in C++. The automatic generation of the real-emission matrix
elements, along with a suitable set of Catani-Seymour \cite{CS} subtraction
terms, have been implemented \cite{AutomatedAmegic} in its
framework. All the phase-space integrations for the results presented
below have been performed using {\tt Sherpa}'s efficient multi-channel
phase-space integration.

{\tt BlackHat} \cite{BlackHatI} is a C++ library
aimed at automating the computation of one-loop matrix elements,
and based on the modern unitarity
method~\cite{UnitarityMethod,Z4Partons}. In general, one-loop matrix
elements can be decomposed into two terms. The first term, called the
``cut part'', contains all the functions that have branch cuts.  The
second ``rational part'' only contains rational functions of spinor
products. The cut part itself can be decomposed at one loop into a sum
of coefficients multiplying scalar one-loop scalar integrals. The
coefficients of these integrals can be determined purely numerically
at the integrand level~\cite{BCFUnitarity,delAguila,OPP} in a
way that meshes well with the unitarity method.  In {\tt BlackHat}, 
a numerical version of Forde's analytic
approach~\cite{BCFUnitarity,Forde} accomplishes this. The rational
part is computed in {\tt BlackHat} using either on-shell recursion
\cite{Bootstrap,BlackHatI} or a variant of $D$-dimensional
unitarity~\cite{DdimUnitarity} along the lines of
ref.~\cite{Badger}.

\section{Tevatron Results}
\begin{figure}
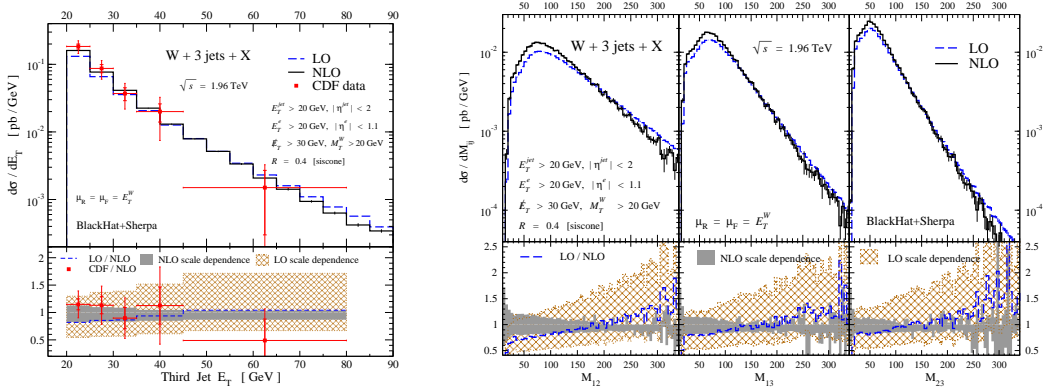

\includegraphics[scale=0.25]{W3jTev_ETWmu_siscone_eb_jets_jet_1_1_Et_3_with_CDF_data.eps}\qquad
\includegraphics[scale=0.35]{W3jTev_ETWmu_siscone_eb_jets_jet_1_1_DJMall.eps}
\caption{Left panel: 
$E_T$ distribution of the third jet at the Tevatron.  Right panel: dijet mass
distributions. In both plots the LO and NLO results are represented
by the blue and black lines, respectively. The (red) data points in the
left panel show the $E_T$ distribution measured by CDF~\cite{WCDF},
including experimental errors. The lower panels show the data and the
LO results normalized by the NLO result.}
\label{fig1}
\end{figure}
The left panel of Figure~\ref{fig1} gives the $E_T$ distribution of the third
jet, and the right panel the dijet mass distributions, for $W\,+\,3$-jet
production at the Tevatron. The former results agree well with data from
CDF~\cite{WCDF}.  The lower panels show the LO and NLO scale variation
bands, and the
data, normalized by the central NLO result, in order to illustrate the shape
difference between LO and NLO.  As expected, the NLO scale variation,
represented by the shaded grey band, is much smaller than the LO
one, represented by the brown hatched bands.  We use the CTEQ6 \cite{CTEQ6M}
PDF sets.  The Tevatron data in the plots have been obtained using
the infrared-unsafe JETCLU jet algorithm \cite{JETCLU}.  We cannot use
this algorithm in our NLO analysis, so we use the SISCONE~\cite{SISCONE}
algorithm instead.

\section{Scale choices}

\begin{figure}
\center\includegraphics[scale=0.4]{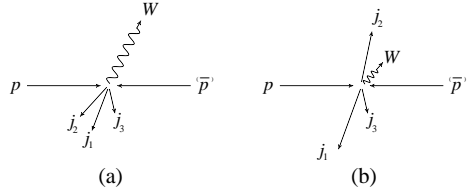}
\caption{Different kinematic configurations for $W+3$ jets:
in (a) the $W$ boson recoils against the three jets; in (b) the $W$
is relatively soft compared to the jets.}
\label{fig2}
\end{figure}

We choose the (common) 
factorization and renormalization scale dynamically,
on an event-by-event basis. The wide range of scales available at
the LHC makes choosing a reasonable scale more important than at the
Tevatron. In multijet
processes, typically more than one scale is involved, which no single
scale choice can fully model.  A bad choice of scale can manifest
itself as a strong dependence of the ratio of NLO to LO cross sections,
or $K$ factor, as a function of the observable considered.  A poorly
chosen scale can even turn the NLO differential cross section negative
in the tails of distributions, because of uncancelled large logarithms
between the chosen scale and the typical scale in the
process~\cite{BergerW3}.

In particular, consider the two configurations shown in Figure~\ref{fig2}
for the case of $W+3$ jets.  In case (a), the three jets recoil against
the $W$ boson, whereas
in case (b) most of the energy is taken by the jets
and the $W$ boson is soft. Choosing the transverse energy of the $W$,
$E_T^W$, as the scale is a good choice for case (a) but rather poor
for case (b), since it does not capture the larger energy scale of the
jets. On the other hand, choosing the total transverse energy, $H_T$
(the scalar sum of the jet, electron and neutrino $E_T$s),
as the scale interpolates better between cases (a) and (b).  In
light of this, our default scale choice for the LHC is a partonic
version of the total transverse energy, $\hat H_T$.

\begin{figure}
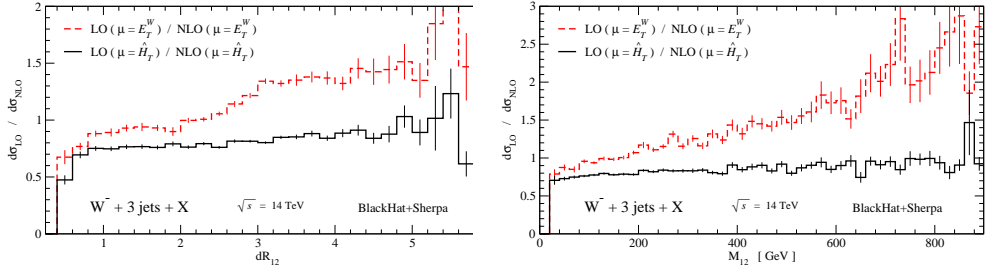

\includegraphics[scale=0.275]{Wm3jLHC_ETW_HT_lo_over_nlo_central_mu_siscone_jets_jet_1_1_dR2__A1.eps}
\includegraphics[scale=0.275]{Wm3jLHC_ETW_HT_lo_over_nlo_central_mu_siscone_jets_jet_1_1_DJM_B1.eps}\quad
\caption{Comparison between two scale choices
for the jet angular separation $\Delta R_{12}$ (left) and dijet mass
$M_{12}$ (right) for the two hardest jets.  The red line represents 
the choice $\mu=M_W^T$ and the black line $\mu=\hat H_T$.}
\label{fig3}
\end{figure}
  
In order to assess the given scale choice, we plot in Figure~\ref{fig3}
the ratio of LO to NLO for the jet angular separation $\Delta R_{12}$ and 
dijet mass $M_{12}$ distributions for the two highest-$E_T$ 
jets in the event, for both the $E_T^W$ and $\hat H_T$ scale choices.
The curve corresponding to the scale $\hat H_T$ is much flatter,
indicating that this choice is better.  A systematic discussion of
scale choices, and their pitfalls, may be found in ref.~\cite{BergerW3}.

\section{LHC results}
\begin{figure}
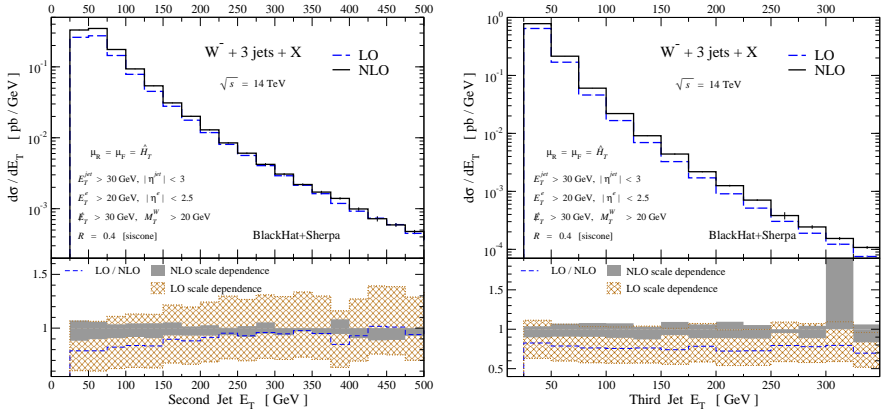

\includegraphics[scale=0.27]{Wm3jLHC_HTmu_siscone_eb_jets_jet_1_1_Et_2.eps}\quad
\includegraphics[scale=0.27]{Wm3jLHC_HTmu_siscone_eb_jets_jet_1_1_Et_3.eps}
\caption{
The left panel shows the $E_T$ distribution of the second hardest
jet at the LHC, while the right panel shows the third jet $E_T$.
In both graphs, the LO and NLO results are represented by the 
blue (dashed) and black lines respectively. The lower panel shows the 
LO and NLO scale variation bands, normalized by the central NLO result. 
}
\label{fig4}
\end{figure}

Figure~\ref{fig4} shows the second and third jet $E_T$ distribution in
$W^-+3$-jet production at the LHC. The lower panels
show the reduced scale dependency of the NLO result as compared to
the LO one. 

\section{Conclusion}

In these proceedings we outlined some sample results from a full NLO
calculation of $W+3$-jet production at the Tevatron and the LHC.
We also summarized
the suitability of the total transverse energy $H_T$ (or a fixed
fraction of it) as a suitable scale choice for observables involving a
$W$ boson in conjunction with many jets. A more detailed discussion,
including many more distributions, may be found in 
ref.~\cite{BergerW3}.
\section*{Acknowledgments}
This research was supported by the US Department of Energy under
contracts DE--FG03--91ER40662, DE--AC02--76SF00515 and
DE--FC02--94ER40818.
DAK's research is supported by the European Research Council under
Advanced Investigator Grant ERC--AdG--228301. HI is supported by a fellowship
from the US LHC Theory Initiative through NSF grant PHY-0705682.
This research used resources of Academic Technology Services at UCLA,
PhenoGrid using the
GridPP infrastructure, and the National Energy Research Scientific
Computing Center, which is supported by the Office of Science of the
U.S. Department of Energy under Contract No. DE-AC02-05CH11231.


\begin{thebibliography}{99}

\bibitem{MCFM}
J.~M.~Campbell and R.~K.~Ellis,
Phys.\ Rev.\  D {\bf 65}, 113007 (2002)
[hep-ph/0202176].

\bibitem{ICHEPBH}
C.~F.~Berger, Z.~Bern, L~.J. Dixon, F.~Febres Cordero,
D.~Forde, H.~Ita, D.~A.~Kosower and D.~Ma\^{\i}tre,
0808.0941 [hep-ph];
%
R.~K.~Ellis, K.~Melnikov and G.~Zanderighi,
JHEP {\bf 0904}, 077 (2009)
[0901.4101 [hep-ph]];
%
R.~K.~Ellis, K.~Melnikov and G.~Zanderighi,
0906.1445 [hep-ph];
%
C.~F.~Berger {\it et al.},
Phys.\ Rev.\ Lett.\ {\bf 102}, 222001 (2009)
[0902.2760 [hep-ph]].

\bibitem{BergerW3}
C.~F.~Berger {\it et al.},
0907.1984 [hep-ph].

\bibitem{SHERPA}
T.~Gleisberg, S.~Hoche, F.~Krauss, M.~Schonherr, S.~Schumann, F.~Siegert and J.~Winter,
JHEP {\bf 0902} (2009) 007
[0811.4622 [hep-ph]].

\bibitem{CS}
S.~Catani and M.~H.~Seymour,
Nucl.\ Phys.\  B {\bf 485}, 291 (1997)
[Erratum-ibid.\  B {\bf 510}, 503 (1998)]
[hep-ph/9605323].

\bibitem{AutomatedAmegic}
T.~Gleisberg and F.~Krauss,
Eur.\ Phys.\ J.\  C {\bf 53}, 501 (2008)
[0709.2881 [hep-ph]].


\bibitem{BlackHatI}
C.~F.~Berger,
Z.~Bern, L.~J.~Dixon, F.~Febres Cordero, D.~Forde, H.~Ita,
D.~A.~Kosower and D.~Ma\^{\i}tre,
Phys.\ Rev.\ D {\bf 78}, 036003 (2008)
[0803.4180 [hep-ph]].

\bibitem{UnitarityMethod}
Z.~Bern, L.~J.~Dixon, D.~C.~Dunbar and D.~A.~Kosower,
Nucl.\ Phys.\  B {\bf 425}, 217 (1994)
[hep-ph/9403226]; 
%
Nucl.\ Phys.\  B {\bf 435}, 59 (1995)
[hep-ph/9409265].

\bibitem{Z4Partons}
Z.~Bern, L.~J.~Dixon and D.~A.~Kosower,
Nucl.\ Phys.\  B {\bf 513}, 3 (1998)
[hep-ph/9708239].


\bibitem{BCFUnitarity}
R.~Britto, F.~Cachazo and B.~Feng,
Nucl.\ Phys.\  B {\bf 725}, 275 (2005)
[hep-th/0412103].

\bibitem{delAguila}
F.~del Aguila and R.~Pittau,
JHEP {\bf 0407} (2004) 017
[hep-ph/0404120].

\bibitem{OPP}
G.~Ossola, C.~G.~Papadopoulos and R.~Pittau,
Nucl.\ Phys.\  B {\bf 763}, 147 (2007)
[hep-ph/0609007].

\bibitem{Forde}
D.~Forde,
Phys.\ Rev.\  D {\bf 75}, 125019 (2007)
[0704.1835 [hep-ph]].

\bibitem{Bootstrap}
Z.~Bern, L.~J.~Dixon and D.~A.~Kosower,
Phys.\ Rev.\  D {\bf 71}, 105013 (2005)
[hep-th/0501240];
Phys.\ Rev.\  D {\bf 72}, 125003 (2005)
[hep-ph/0505055];
Phys.\ Rev.\  D {\bf 73}, 065013 (2006)
[hep-ph/0507005];
%
D.~Forde and D.~A.~Kosower,
Phys.\ Rev.\  D {\bf 73}, 065007 (2006)
[hep-th/0507292];
%
Phys.\ Rev.\  D {\bf 73}, 061701 (2006)
[hep-ph/0509358];
%
C.~F.~Berger, Z.~Bern, L.~J.~Dixon, D.~Forde and D.~A.~Kosower,
Phys.\ Rev.\  D {\bf 74} (2006) 036009
[arXiv:hep-ph/0604195].

\bibitem{DdimUnitarity}
Z.~Bern and A.~G.~Morgan,
Nucl.\ Phys.\  B {\bf 467}, 479 (1996)
[hep-ph/9511336];
Z.~Bern, L.~J.~Dixon, D.~C.~Dunbar and D.~A.~Kosower,
Phys.\ Lett.\  B {\bf 394}, 105 (1997)
[hep-th/9611127];
%
C.~Anastasiou, R.~Britto, B.~Feng, Z.~Kunszt and P.~Mastrolia,
Phys.\ Lett.\  B {\bf 645}, 213 (2007)
[hep-ph/0609191];
%
R.~Britto and B.~Feng,
JHEP {\bf 0802}, 095 (2008)
[0711.4284 [hep-ph]];
%
W.~T.~Giele, Z.~Kunszt and K.~Melnikov,
  JHEP {\bf 0804} (2008) 049
  [arXiv:0801.2237 [hep-ph]];
R.~K.~Ellis, W.~T.~Giele, Z.~Kunszt and K.~Melnikov,
0806.3467 [hep-ph].

\bibitem{Badger}
S.~D.~Badger,
JHEP {\bf 0901}, 049 (2009)
[0806.4600 [hep-ph]].

\bibitem{WCDF}
T.~Aaltonen {\it et al.}  [CDF Collaboration],
Phys.\ Rev.\  D {\bf 77}, 011108 (2008)
[0711.4044 [hep-ex]].

\bibitem{CTEQ6M}
J.~Pumplin {\it et al.},
JHEP {\bf 0207}, 012 (2002)
[hep-ph/0201195].

\bibitem{JETCLU}
F.~Abe {\it et al.}  [CDF Collaboration],
Phys.\ Rev.\  D {\bf 45}, 1448 (1992).

\bibitem{SISCONE}
G.~P.~Salam and G.~Soyez,
JHEP {\bf 0705}, 086 (2007)
[0704.0292 [hep-ph]].

\end{thebibliography}
\end{document}